\documentclass[prl,preprintnumbers,amsmath,amssymb,tightenlines,twocolumn,superscriptaddress]{revtex4}

\usepackage{graphicx}% Include figure files
\usepackage{dcolumn}% Align table columns on decimal point
\usepackage{bm}% bold math
\usepackage{epsfig}
\newcommand{\bra}[1]{\langle\,{#1}\, |}
\newcommand{\ket}[1]{|\,{#1}\,\rangle}

\def\beq{\begin{equation}}
\def\eeq{\end{equation}}

\newcommand{\fref}[1]{Fig.~\ref{#1}}

\newcommand{\eref}[1]{Eq.~(\ref{#1})}

\newcommand{\cref}[1]{chapter~\ref{#1}}

\newcommand{\Cref}[1]{Chapter~\ref{#1}}

\newcommand{\dipole}{J}

\usepackage{ulem}  %unter- (\uline{important}) und durchstreichen (\sout{wrong})
\usepackage{color}

\newcommand{\oac}[1]{}
\begin{document}

\title{Tuning nonradiative lifetimes via molecular aggregation}

\author{A.~Celestino}
\affiliation{Max Planck Institute for the Physics of Complex Systems, N\"othnitzer Strasse 38, 01187 Dresden, Germany}
\author{A.~Eisfeld}
\affiliation{Max Planck Institute for the Physics of Complex Systems, N\"othnitzer Strasse 38, 01187 Dresden, Germany}

\email{eisfeld@pks.mpg.de}
\begin{abstract}
We show that molecular aggregation can strongly influence the nonradiative decay (NRD) lifetime of an electronic excitation. As a demonstrative example, we consider a transition-dipole-dipole-interacting dimer whose monomers have harmonic potential energy surfaces (PESs). Depending on the position of the NRD channel ($q_{\rm nr}$), we find that the NRD lifetime ($\tau_{\rm nr}^{\rm dim}$) can exhibit a completely different dependence on the intermolecular-interaction strength. We observe that (i) for $q_{\rm nr}$ near the Franck-Condon region, $\tau_{\rm nr}^{\rm dim}$ increases with the interaction strength; (ii) for $q_{\rm nr}$ near the minimum of the monomer excited PES, the intermolecular interaction has little influence on $\tau_{\rm nr}^{\rm dim}$; (iii) for $q_{\rm nr}$ near the classical turning point of the monomer nuclear dynamics, on the other side of the minimum, $\tau_{\rm nr}^{\rm dim}$ decreases with the interaction strength. Our findings suggest design principles for molecular systems where a specific fluorescence quantum yield is desired.

\end{abstract}
%
%\pacs{						%| 
%82.20.Rp,  % State to state energy transfer 			%| -> these three are as in the cradle paper
%34.20.Cf,   % Interatomic potentials and forces			%| -> I think this is wrong. One could use instead:
%34.20.Gj	,   % Intermolecular and atom-molecule potentials and forces -> New
%33.50.Hv,   %	Radiationless transitions, quenching -> New
%33.70.Ca,   %	Oscillator and band strengths, lifetimes, transition moments, and Franck-Condon factors -> New, but maybe shouldnt enter
%31.70.Hq	 % Time-dependent phenomena: excitation and relaxation processes, and reaction rates -> New
%}

\maketitle

\begin{figure}[pt]
\includegraphics[width=8cm]{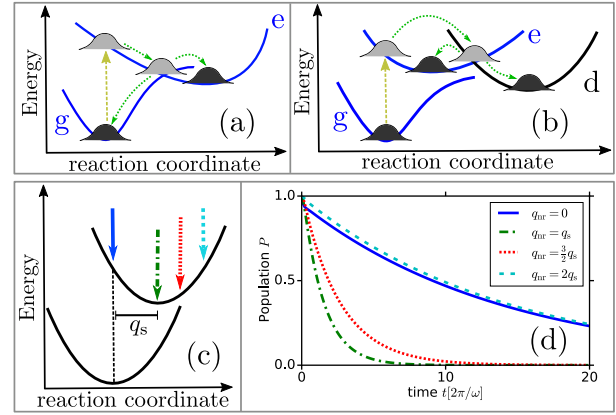}
\caption{\label{fig:sketch_monomer}
(Color online) Potential energy surfaces (PESs) and nonradiative decay (NRD) dynamics of a single molecule. The top row shows sketches of the electronic excitation and relaxation processes along the PESs. After a vertical Franck-Condon transition, the electronic relaxation can occur in two different ways: (a) direct relaxation from the electronically excited state e to the electronic ground state g; (b) relaxation from the optically bright state e to the electronic dark state d. (c) Ground and first optically excited state harmonic PESs (case considered in the numerics). $q_{\rm s}$ is the shift between PESs. (d) Time-resolved population decay for different positions of the NRD channel. The $q_{\rm nr}$ values are illustrated as arrows in (c) according to the colors and linestyles of the curves in (d).
}
\end{figure}

%%%%%%%%%%%%%%%%%%%%%%%%%%%%%%%%%%%
%%%%%%%%%%%%%%%%%%%%%%%%%%%%%%%%%%%
%\ssection{Introduction}
%%%%%%%%%%%%%%%%%%%%%%%%%%%%%%%%%%%
%%%%%%%%%%%%%%%%%%%%%%%%%%%%%%%%%%%
The lifetime of a molecular system's electronically excited state (EES) is determined by radiative and nonradiative transitions \cite{Si83__,MeOs95__,La06__,MaKue11__}. While radiative electronic transitions in molecular systems commonly occur within nanoseconds, nonradiative transitions can be much faster, with picosecond or even femtosecond timescales. Thus nonradiative decay (NRD) processes can determine the EES lifetime in molecular systems. Often these fast NRD processes are useful, e.g.~in switching between retinal isomers \cite{PoAlWe10_440_,HaSt00_1146_,HaSt00_297_,ScLiKu15_2886_}, transfer to long-lived triplet states \cite{LiPaYa14_892_}, excitation quenching in photosynthesis \cite{NiBjGr97_14162_,RuYoHo96_674_,BrShGh15__}, DNA photoprotection \cite{MiLaSu09_217_,BaAqSz10_21453_,ChWeRo14_843_}, singlet fission \cite{MuLiSc15_352_}, molecular rotary motors \cite{KaKiSc10_5058_}, and molecular switches \cite{QuDoGe14_8756_,LeScBr12_15296_,JiXiLi11_244_}.
However, sometimes these NRD processes are unwanted, e.g.~in light harvesting applications \cite{FaGrMa12_6367_} or where large emission rates are desirable \cite{XiChQu11_926_,DuHoLe10_6392_}.

In this Letter we show that molecular aggregation can strongly influence the timescale of NRD processes (NRD lifetime) in molecular systems. Molecular aggregates consisting of transition-dipole-dipole-interacting molecules have attracted interest for decades (see e.g.\ Refs.~\cite{Ko96__,KueLo11_47_,SaEiVa13_21_} and references therein).
One reason is the significant changes of the optical properties (e.g.~radiative lifetime) upon aggregation, caused by the formation of exciton states coherently delocalized over several molecules. As we will show below, the excitonic delocalization can also strongly influence the NRD lifetime.

The basic features of the single molecule (monomer) are sketched in Fig.~\ref{fig:sketch_monomer} (a) and (b), where the relevant potential energy surfaces (PESs) are shown as a function of a single nuclear ``reaction'' coordinate $q$.
Initially the molecule is in its electronic ground state $\ket{\rm g}$, in thermal equilibrium with respect to the ground state PES.
After a vertical Franck-Condon transition (e.g.~through a short laser pulse) to an EES $\ket{\rm e}$, which leaves the nuclear wavefunction unchanged \cite{La06__}, the nuclear dynamics and the NRD set in.
We denote the PESs of the electronic ground and excited states by $V_{\rm g}(\vec{Q})$ and $V_{\rm e}(\vec{Q})$, respectively.
Here, $\vec{Q}$ is the set of all relevant nuclear coordinates. Vibrational relaxation due to coupling to environmental degrees of freedom accompanies the coherent motion on the PESs.

Typically, nonradiative transitions between molecular electronic states involve nuclear degrees of freedom and occur at points where the respective PESs are close or cross \cite{Si83__,DoYaKoe04__}. We assume a localized region in nuclear space where the electronic excitation can efficiently leave the electronic state $\ket{\rm e}$, which we call ``the NRD channel''.  Since we do not focus on a particular molecule and we are mainly interested in qualitative results, we model the NRD channel as an imaginary potential added to the excited PES. This PES is then given by
$\tilde{V}_{\rm e}(\vec{Q})=V_{\rm e}(\vec{Q})-i\Gamma(\vec{Q})$, where we denote $\Gamma(\vec{Q})$ as the ``decay-function''. Note that this imaginary potential implies that the Hamiltonian of the system is non-Hermitian.
We emphasize that this NRD channel can occur in any region of the PES. For instance, in $\beta$-Apo-$8$'-carotenal a NRD channel is believed to occur at the vertical Franck-Condon region \cite{OlFl15_11428_}.
  
As a concrete example, we consider harmonic monomer PESs along a single normal mode coordinate \cite{MaKu11__} (reaction coordinate $q$), with identical frequencies $\omega$ but shifted with respect to each other by $q_{\rm s}$ (see Fig.~\ref{fig:sketch_monomer} (c)). The monomer PESs are $V_{\rm g}(q)=\frac{1}{2}\omega^2q^2$ and $V_{\rm e}(q)=E_{\rm e}+\frac{1}{2}\omega^2(q-q_{\rm s})^2$.
Here, $E_{\rm e}$ is the energy shift between the minima of the PESs (electronic transition energy). For simplicity, we take the decay-function to be a delta function $\Gamma(q)=\lambda\delta (q-q_{\rm nr})$ centered at the position $q_{\rm nr}$, where $\lambda$ is the NRD strength. Vibrational relaxation is taken into account via a bilinear coupling between the system and a bath of harmonic oscillators. The system part of the coupling operator is proportional to $\left(q-q_{\rm s}\ket{\rm e}\bra{\rm e}\right)$. We have used the Ohmic spectral density $j\left(\tilde{\omega}\right)=\theta\left(\tilde{\omega}\right)\gamma\tilde{\omega} \exp\left(-\tilde{\omega}/\omega_0\right)$ ($\theta(\tilde{\omega})$ is the Heaviside step function). We simulate the system's dynamics solving a multilevel Redfield equation \cite{MaKu11__}, obtained after tracing out the bath degrees of freedom. For more details on the modeling of vibrational relaxation, we refer to the Supplemental Material. All numerical results shown in this Letter are obtained for the parameter values $q_{\rm s}=1.5\sqrt{\hbar/\omega}$, $\gamma=\hbar^2/\pi$, $\omega_0=10\omega/\pi$, NRD strength $\lambda=0.1\hbar^{3/2}\omega^{1/2}$, and temperature $T=0$ (since we are not interested in thermal effects). The values of $\gamma$, $\omega_0$, and $\lambda$ guarantee fast vibrational relaxation compared to the timescales of the NRD and to nuclear oscillations. This case applies to many molecules, as vibrational relaxation typically takes place within a picosecond \cite{La06__}.

As a reference for the dimer case later on, we now consider the dependence of the NRD on $q_{\rm nr}$ for a single molecule. We focus on different locations $q_{\rm nr}$ of the NRD channel leading to qualitatively different behaviors. These locations are $q_{\rm nr}=0$ (at the vertical Franck-Condon region), $q_{\rm nr}=q_{\rm s}$ (minimum of the excited-state PES of the monomer), and $q_{\rm nr}=2q_{\rm s}$ (classical turning point to the right of this minimum). Note that $q_{\rm nr}=0$ and $q_{\rm nr}=2q_{\rm s}$ enclose the classically accessible region in the monomer nuclear space. The numerically calculated population in the monomer EES $P(t)$ is shown in Fig.~\ref{fig:sketch_monomer} (d). Different NRD channel positions $q_{\rm nr}$ are indicated by arrows in Fig.~\ref{fig:sketch_monomer} (c) according to the colors and linestyles in Fig.~\ref{fig:sketch_monomer} (d). $P(t)$ depends sensitively on $q_{\rm nr}$, and decays approximately as a monoexponential $P(t)\approx \exp\left(-t / \tau_{\rm nr}^{\rm (mon)}\right)$ because vibrational relaxation is fast compared to NRD dynamics. We estimate the monomeric NRD lifetime assuming that the nuclear wavefunction is in the ground state of $V_{\rm e}(q)$ at all times (with a different, time-dependent norm). The obtained expression $\tau_{\rm nr}^{\rm (mon)}(q_{\rm nr})\approx\tau_{\rm nr}^{\rm (mon)}(q_{\rm s})\exp[\omega(q_{\rm nr}-q_{\rm s})^2/\hbar]$ ($\tau_{\rm nr}^{\rm (mon)}(q_{\rm s})\approx\sqrt{\pi}\hbar^{3/2}/2\lambda\omega^{1/2}$) fits well the numerical results from Fig.~\ref{fig:sketch_monomer} (d). In particular, the closer the NRD channel is to the minimum of the excited-state PES, the faster the NRD takes place.

\begin{figure}
\includegraphics[width=7cm]{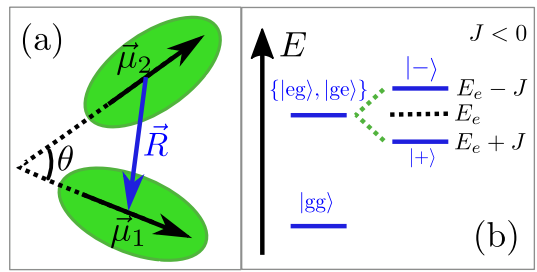}
\caption{\label{fig:Dimer_skizze}(Color online) Sketch of the geometry (a) and energy levels (b) (vibronic levels suppressed) of the dimer. (a) The dimer's geometry is defined by the relative orientation of
the transition dipoles $\vec{\mu}_{1}$ and $\vec{\mu}_{2}$ ($\theta$), and the relative position $\vec{R}$. (b) The eigenenergies of the ground state $\ket{\rm gg}$, single excitation manifold $\{\ket{\rm eg},\ket{\rm ge}\}$ for vanishing $J$ ($E_{\rm e}$), and dimer singly excited states $|+\rangle$ and $|-\rangle$ (respectively $E_{\rm e}+J$ and $E_{\rm e}-J$) for $J<0$.}
\end{figure}

\begin{figure}[tbp]
\includegraphics[width=8cm]{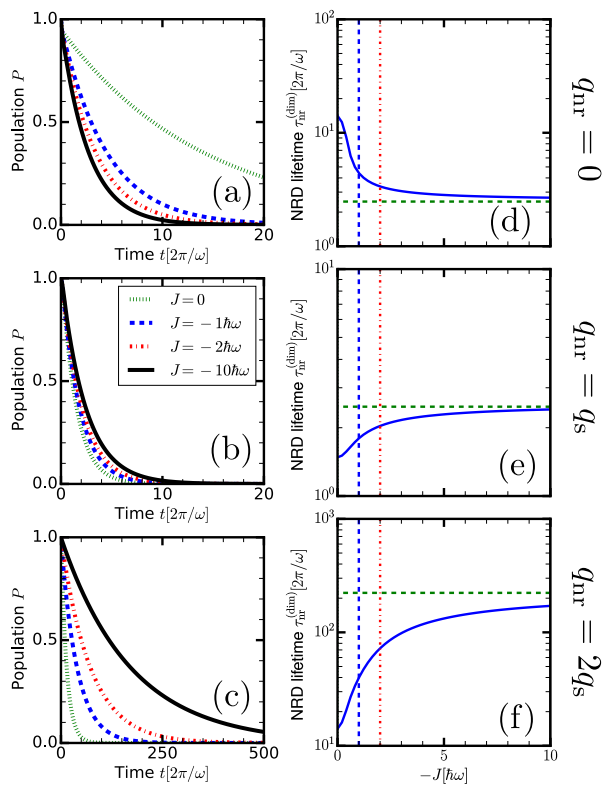}
\caption{
(Color online) Nonradiative decay dynamics for different $J$ and $q_{\rm nr}$ values. (a), (b), and (c): Population in the single excitation manifold $P(t)$ as a function of time. (d), (e), and (f): NRD lifetime $\tau_{\rm nr}^{\rm (dim)}(J)$ as a function of $J$ (continuous blue line). The green dashed horizontal line is the NRD lifetime saturation value $\tau_{\rm nr}^{\rm (sat)}$. For clarity, the values $J=-1$ and $J=-2$ for which we plotted $P(t)$ in (a), (b), and (c) are indicated by vertical lines matching the colors and linestyles in (a), (b), and (c).}

\label{gmaXqnr}
\end{figure}

To furnish a clear example on how the transition dipole-dipole interaction influences the NRD lifetime, we treat the case of a molecular dimer \cite{WiMo60_872_,FuGo61_1059_,Me63_154_,FuGo64_2280_,BeDaKj02_5810_,RoEiDv11_054907_} in detail (see Fig.~\ref{fig:Dimer_skizze}). The two monomers are assumed to be sufficiently far apart to neglect overlap between electronic wavefunctions.
However, they interact via long-range Coulomb interaction. The Coulomb interaction depends on the dimer's geometry, which is considered fixed (see Fig.~\ref{fig:Dimer_skizze} (a)). In the point-dipole approximation, which is often appropriate, the interaction strength can be written as $J\propto\frac{1}{R^3}\left(\vec{\mu}_1\cdot\vec{\mu}_2 - 3 (\vec{R}\cdot\vec{\mu}_1)(\vec{R}\cdot\vec{\mu}_2)/R^2\right)$. Here, $\vec{R}$ is the distance vector between the centers of the two monomers and we consider $\vec{\mu}_1$ and $\vec{\mu}_2$ to be the transition dipoles of monomer 1 and 2, respectively.
We stress that the specific form of this interaction is not relevant in the following.

The electronic subspace is spanned by the states $\ket{\rm gg}$,  $\ket{\rm eg}$,
 $\ket{\rm ge}$, and $\ket{\rm ee}$ (see Fig.~\ref{fig:Dimer_skizze} (b)).
For both monomers in the electronic ground state, the corresponding nuclear Hamiltonian is $H_{\rm gg}(\vec{Q}_1,\vec{Q}_2)=H_{\rm g}(\vec{Q}_1)+H_{\rm g}(\vec{Q}_2)$, where $H_{\rm g}(\vec{Q}_j)=K_j+V_{\rm g}(\vec{Q}_j)$, and $K_j$ is the nuclear kinetic energy for the monomer $j$.
Consequently, the initial state (before the Franck-Condon vertical transition) is the same as the thermal equilibrium of two uncoupled monomers.
Because of large detuning in energy, the doubly excited state $\ket{ee}$ is not populated and we will not discuss it further.
In the single excitation manifold, i.e.~in the subspace spanned by the degenerate electronic states $\ket{\rm eg}$ and $\ket{\rm ge}$, the transition dipole-dipole interaction leads to a coupling of the form $J \left(\ket{\rm eg}\bra{\rm ge}+ \ket{\rm ge}\bra{\rm eg}\right)$. The nuclear Hamiltonian in the single excitation manifold is then given by
\begin{align}
\label{eq:H_ex_mit_vib}
H_{\rm ex}\left(\vec{Q}_1,\vec{Q}_2\right)=K_{\rm nuc}&+\left(\tilde{V}_{\rm e}\left(\vec{Q}_1\right)+V_{\rm g}\left(\vec{Q}_2\right)\right)\ket{\rm eg}\bra{\rm eg} \nonumber\\ 
&+\left(\tilde{V}_{\rm e}\left(\vec{Q}_2\right)+V_{\rm g}\left(\vec{Q}_1\right)\right)\ket{\rm ge}\bra{\rm ge}\nonumber\\
&+ J \left(\ket{\rm eg}\bra{\rm ge}+ \ket{\rm ge}\bra{\rm eg}\right),
\end{align}
where $K_{\rm nuc}=K_1+K_2$ and we note that $H_{\rm ex}$ is non-Hermitian.

%%%%%%%%%%%%%%%%%%%%%%%%%%%%%%%%%%%
%%%%%%%%%%%%%%%%%%%%%%%%%%%%%%%%%%%
%\ssection{Numerical simulations}
%%%%%%%%%%%%%%%%%%%%%%%%%%%%%%%%%%%
%%%%%%%%%%%%%%%%%%%%%%%%%%%%%%%%%%%
The monomers have the same PESs and NRD channels as the single molecule considered when discussing \fref{fig:sketch_monomer} (d). This model directly relates to previous studies of dimers, where NRD has not been taken into account (see e.g.\ Refs.~\cite{WiMo60_872_,FuGo61_1059_,Me63_154_,FuGo64_2280_,BeDaKj02_5810_,RoEiDv11_054907_}). Each monomer is bilinearly coupled to a distinct harmonic bath. The system part of the coupling operator is now proportional to $\left(q_j-q_{\rm s}\ket{\pi_j}\bra{\pi_j}\right)$, where $\ket{\pi_1}=\ket{eg}$ and $\ket{\pi_2}=\ket{ge}$. As in the monomer case, we derive a multilevel Redfield master equation for the dimer by tracing out the environmental degrees of freedom \cite{MaKu11__} and performing Born, Markov, and secular approximations. This equation is
\begin{equation}
\label{eq:master}
\frac{\partial \rho}{\partial t}=-\frac{i}{\hbar}\left(H_{\rm ex}\rho-\rho H_{\rm ex}^\dag\right)+\mathcal{L}\left[\rho\right],
\end{equation}
where $t$ is the time, $\rho$ is the density matrix of the dimer, and $\mathcal{L}$ is a dissipator. For more details on the modeling of the vibrational relaxation, we refer to the Supplemental Material. As our initial state, we consider the result of a Franck-Condon transition to the first excited adiabatic electronic state. For $J<0$, the case we show here, this corresponds to $|+\rangle=(\ket{eg}+\ket{ge})/\sqrt{2}$ (for $J>0$ it corresponds to $|-\rangle=(\ket{eg}-\ket{ge})/\sqrt{2}$). These two electronic states are the eigenstates of the system if nuclear degrees of freedom are neglected (see Fig.~\ref{fig:Dimer_skizze} (b)). Note that the results we discuss here are not fundamentally changed by choosing a different initial condition within the single excitation manifold. We use the same values of $q_{\rm s}$, $\gamma$, $\omega_0$, $\lambda$, and $T$, as in the monomer case discussed above. We also stress that the baths are at the same temperature and have the same spectral density.

The numerical results for the dimer are shown in \fref{gmaXqnr}. From the top to the bottom row, $q_{\rm nr}=0$, $q_{\rm nr}=q_{\rm s}$, and $q_{\rm nr}=2q_{\rm s}$ are shown, respectively. In the left column, the population in the single excitation manifold $P(t)$ is shown for different values of $J$. As in the monomer case, $P(t)$ approximately follows a monoexponential decay (see \fref{gmaXqnr} (a)-(c)) and can therefore be written as $P(t)\approx \exp\left(-t / \tau_{\rm nr}^{\rm (dim)}\right)$. The numerically fitted NRD lifetime $\tau_{\rm nr}^{\rm (dim)}$ is plotted as a function of $J$ as continuous blue lines in the right column (\fref{gmaXqnr} (d)-(f)). As one can see from \fref{gmaXqnr} (d)-(f), $P(t)$ depends on $J$, and this dependence is different for different values of $q_{\rm nr}$. For all $q_{\rm nr}$, the NRD lifetime $\tau_{\rm nr}^{\rm (dim)}$ varies monotonically with $J$ and it saturates for small $J$ ($J<0$) at the value $\tau_{\rm nr}^{\rm (sat)}$, which depends on the specific choice of $q_{\rm nr}$, $q_{\rm s}$ and $\lambda$. This saturation value can be analytically determined to be $\tau_{\rm nr}^{\rm (sat)}(q_{\rm nr})\approx\tau_{\rm nr}^{\rm (sat)}(q_{\rm s}/2)\exp[\omega(q_{\rm nr}-q_{\rm s}/2)^2/\hbar]$ (see discussion about the adiabatic limit below), with $\tau_{\rm nr}^{\rm (sat)}(q_{\rm s}/2)\approx\pi^{1/2}\hbar^{3/2}/2\lambda\omega^{1/2}$, and is plotted as a dashed green line in \fref{gmaXqnr} (d)-(f). The value of $q_{\rm nr}$ determines whether $\tau_{\rm nr}^{\rm (dim)}$ increases or decreases with $J$. For $q_{\rm nr}>3q_{\rm s}/4$ ($q_{\rm nr}<3q_{\rm s}/4$), $\tau_{\rm nr}^{\rm (dim)}$ increases (decreases) with $|J|$.

The transition dipole-dipole interaction can suppress (trigger) fluorescence of (non-)fluorescent molecules when they form dimers or larger aggregates. Although in the examples shown in \fref{gmaXqnr} (d)-(f) the NRD lifetime maximally varied over approximately one order of magnitude (\fref{gmaXqnr} (f)), this is not limited on the range of NRD lifetime variation. The range of variation of $\tau_{\rm nr}^{(dim)}$ with $J$ is at least the ratio $\tau_{\rm nr}^{\rm (mon)}/\tau_{\rm nr}^{\rm (sat)}\approx\exp\left[\omega\left(3q_{\rm s}^2/4-q_{\rm nr}q_{\rm s}\right)/\hbar\right]$. Since it depends exponentially on the shift between the monomer PESs $q_{\rm s}$, the range of tunability becomes exponentially larger for larger $q_{\rm s}$.

The analytic expression of $\tau_{\rm nr}^{\rm (sat)}$ we used in the discussion of our numerical results was derived in the adiabatic limit. In this limit, namely when $|J| \gg \hbar^{1/2}\omega^{3/2}q_{\rm s}/\sqrt{2}$, we can consider the nuclear dynamics to be confined within the adiabatic PES associated with the electronic state $|+\rangle$ ($|-\rangle$) for $J<0$ ($J>0$). The corresponding (complex) PESs are given by
\begin{equation}
\label{eq:Vpm}
\tilde{V}_{\pm}(\vec{Q}_1,\vec{Q}_2)=\sum_{j=1}^2\frac{1}{2}\big(\tilde{V}_{\rm e}(\vec{Q}_j)+V_{\rm g}(\vec{Q}_j)\big)\pm \dipole,
\end{equation}
and non-adiabatic couplings between these PESs are negligible. Notice from Eq.~(\ref{eq:Vpm}) that the coordinates $\vec{Q}_1$ and $\vec{Q}_2$ are not coupled.
Thus, for each coordinate the NRD channel (which appears via $\tilde{V}_{\rm e}(\vec{Q}_j)$) is the same as for the uncoupled monomers.
However, the potential on which the nuclear wavepacket moves has a different shape (and in particular a different minimum) from the monomer's excited-state PES.
Considering the PESs from our numerics, we obtain for $V_{\pm}(q_1,q_2)$ (the Hermitian part of $\tilde{V}_{\pm}(q_1,q_2)$) a well known result \cite{HaShCh99_381_,BeDaKj02_5810_,EiBrSt05_134103_}: $V_{\pm}(q_1,q_2)=\pm J + \sum_j \omega^2(q_j-q_{\rm s}/2)^2/2 +\omega^2q_{\rm s}^2/4$. The adiabatic PESs are thus shifted by $q_{\rm s}/2$ in each coordinate with respect to the ground-state PES of the monomer. Taking these PESs into account, and considering fast vibrational relaxation compared to the NRD and nuclear oscillations timescale, we obtained the analytic formula for the saturated NRD lifetime $\tau_{\rm nr}^{\rm (sat)}$. This was performed in the same way as for deriving $\tau_{\rm nr}^{\rm (mon)}$, but assuming the nuclear wavefunction to be always in the ground state of $V_{+}(q_1,q_2)$. Comparing the analytic formulas for $\tau_{\rm nr}^{\rm(sat)}$ with $\tau_{\rm nr}^{\rm(mon)}$, one observes that their formulas only differ by the shift in the nuclear coordinate, $q_{\rm s}$ for the monomer and $q_{\rm s}/2$ for the saturated dimer. This is because the minimum of $V_{+}(q_1,q_2)$ is at $(q_1=q_{\rm s}/2,q_2=q_{\rm s}/2)$, while the minimum from the excited-state PES of the monomer $j$ lies at $q_j=q_{\rm s}$.

An extension of the NRD lifetime analysis to longer aggregates can also be performed \cite{ScFi84_269_,WaEiBr08_044505_}. The adiabatic PESs of an $N$-mer (when the electronic wavefunction is delocalized over $N$ monomers) can be estimated in the adiabatic limit. For harmonic monomer PESs, the adiabatic PESs are harmonic and shifted in all reaction coordinates by $q_{\rm s}/N$ (they are also shifted in energy). The ratio between monomeric and saturated N-meric NRD lifetimes behave as $\tau_{\rm nr}^{\rm(mon)}/\tau_{\rm nr}^{{\rm(sat},N)}\sim\exp\left[-\omega q_{\rm s}\left(q_{\rm nr}(2N-2)/N-q_{\rm s}\left(N^2-1\right)/N^2\right)/\hbar\right]$.

%%%%%%%%%%%%%%%%%%%%%%%%%%%%%%%%%%%
%%%%%%%%%%%%%%%%%%%%%%%%%%%%%%%%%%%
%\ssection{Conclusions and Outlook}
%%%%%%%%%%%%%%%%%%%%%%%%%%%%%%%%%%%
%%%%%%%%%%%%%%%%%%%%%%%%%%%%%%%%%%%
In conclusion, we have shown that molecular aggregation can modify the NRD lifetime. As a proof of concept, we have considered in detail the simplest molecular aggregate featuring this phenomenon: a transition-dipole-dipole-interacting dimer with harmonic monomer PESs. We have shown that the relationship between the NRD lifetime and the intermolecular-interaction strength depends sensitively on the NRD channel position. In particular the NRD lifetime can increase with, decrease with, or be practically insensitive to the intermolecular-interaction strength. This indicates that quantum yield measurements can, e.g., be exploited for the detection of molecular aggregation; pinpointing of NRD channel locations in molecules; or to infer the geometry of molecular aggregates. We have also performed simulations for other values of the shift between monomer PESs. If $q_{\rm s}$ is larger than in \fref{gmaXqnr}, e.g.~$q_{\rm s}=3.5\sqrt{\hbar/\omega}$, $\tau_{\rm nr}^{\rm (dim)}$ can feature dips (and peaks) at certain $J$ values - apart from spanning many (e.g.~4 for $q_{\rm nr}=2q_{\rm s}$) orders of magnitude upon varying $J$.

For arbitrary (nonharmonic) monomer PESs, the transition dipole-dipole interaction can impose more severe modifications to the NRD dynamics. This is because the shape of the dimer PESs from \eref{eq:Vpm} can differ from the monomer excited-state PES's shape. For instance, the dimer's PESs may present a minimum even if the excited-state PES of the monomer does not. This can lead to a fundamental change of the nuclear dynamics in the diabatic case, e.g.~stabilizing a photodissociation. Monomer PESs of different shapes will thus give rise to different changes in the NRD dynamics upon aggregation.

%%%%%%%%%%%%%%%%%%%%%%%%%%%%%%%%%%%%%%%%%%%%%%%%%%%%%%%%%%%%%%%%
\acknowledgments

We thank C. Bentley for useful discussions.

%%%%%%%%%%%%%%%%%%%%%%%%%%%%%%%%%%%%%%%%%%%%%%%%%%%%%%%%%%%%%%%%%%%%%%%%%%%
\bibliographystyle{journal_v5}
%\bibliography{./Supression_Relax.bib,/home/eisfeld/01_arbeit/Literatur/bib-dateien/PaperBib.bib,./refs.bib,/home/alan/Desktop/papers/references.bib}
\bibliography{./Supression_Relax,/home/alan/Desktop/papers/references,./refs}

\begin{thebibliography}{10}
\providecommand{\url}[1]{\texttt{#1}}
\providecommand{\urlprefix}{URL }
\expandafter\ifx\csname urlstyle\endcsname\relax
  \providecommand{\doi}[1]{doi:\discretionary{}{}{}#1}\else
  \providecommand{\doi}{doi:\discretionary{}{}{}\begingroup
  \urlstyle{rm}\Url}\fi
\providecommand{\eprint}[2][]{\url{#2}}

\bibitem{Si83__}
J.~Simons; \emph{Energetic Principles of Chemical Reactions}; Jones \& Bartlett
  Pub. (1983).

\bibitem{MeOs95__}
E.~S. Medvedev and V.~I. Osherov; \emph{{Radiationless Transitions in
  Polyatomic Molecules}}; volume~57 of \emph{Springer Series in Chemical
  Physics}; Springer-Verlag (1995).

\bibitem{La06__}
J.~R. Lakowicz; \emph{Principles of Fluorescence Spectroscopy}; Springer Verlag
  (2006).

\bibitem{MaKue11__}
V.~May and O.~K{\"u}hn; \emph{{Charge and Energy Transfer Dynamics in Molecular
  Systems}}; WILEY-VCH; 3rd edition edition (2011).

\bibitem{PoAlWe10_440_}
D.~Polli, P.~Altoe, O.~Weingart, K.~M. Spillane, C.~Manzoni, D.~Brida,
  G.~Tomasello, G.~Orlandi, P.~Kukura, R.~A. Mathies, M.~Garavelli and
  G.~Cerullo; Nature \textbf{467} 440 (2010).

\bibitem{HaSt00_1146_}
S.~Hahn and G.~Stock; The Journal of Physical Chemistry B \textbf{104} 1146
  (2000).

\bibitem{HaSt00_297_}
S.~Hahn and G.~Stock; Chemical Physics \textbf{259} 297  (2000).

\bibitem{ScLiKu15_2886_}
C.~Schnedermann, M.~Liebel and P.~Kukura; Journal of the American Chemical
  Society \textbf{137} 2886 (2015).

\bibitem{LiPaYa14_892_}
W.~Li, Y.~Pan, L.~Yao, H.~Liu, S.~Zhang, C.~Wang, F.~Shen, P.~Lu, B.~Yang and
  Y.~Ma; Advanced Optical Materials \textbf{2} 892 (2014).

\bibitem{NiBjGr97_14162_}
K.~K. Niyogi, O.~Björkman and A.~R. Grossman; Proceedings of the National
  Academy of Sciences \textbf{94} 14162 (1997).

\bibitem{RuYoHo96_674_}
A.~V. Ruban, A.~J. Young and P.~Horton; Biochemistry \textbf{35} 674 (1996).

\bibitem{BrShGh15__}
W.~P. Bricker, P.~M. Shenai, A.~Ghosh, Z.~Liu, M.~G.~M. Enriquez, P.~H.
  Lambrev, H.-S. Tan, C.~S. Lo, S.~Tretiak, S.~Fernandez-Alberti \emph{et~al.};
  Scientific reports \textbf{5} (2015).

\bibitem{MiLaSu09_217_}
C.~T. Middleton, K.~de~La~Harpe, C.~Su, Y.~K. Law, C.~E. Crespo-Hernández and
  B.~Kohler; Annual Review of Physical Chemistry \textbf{60} 217 (2009).

\bibitem{BaAqSz10_21453_}
M.~Barbatti, A.~J.~A. Aquino, J.~J. Szymczak, D.~Nachtigallová, P.~Hobza and
  H.~Lischka; Proceedings of the National Academy of Sciences \textbf{107}
  21453 (2010).

\bibitem{ChWeRo14_843_}
A.~S. Chatterley, C.~W. West, G.~M. Roberts, V.~G. Stavros and J.~R.~R. Verlet;
  The Journal of Physical Chemistry Letters \textbf{5} 843 (2014).

\bibitem{MuLiSc15_352_}
A.~J. Musser, M.~Liebel, C.~Schnedermann, T.~Wende, T.~B. Kehoe, A.~Rao and
  P.~Kukura; Nature Physics \textbf{11} 352 (2015).

\bibitem{KaKiSc10_5058_}
A.~Kazaryan, J.~C.~M. Kistemaker, L.~V. Schäfer, W.~R. Browne, B.~L. Feringa
  and M.~Filatov; The Journal of Physical Chemistry A \textbf{114} 5058 (2010).

\bibitem{QuDoGe14_8756_}
M.~Quick, A.~L. Dobryakov, M.~Gerecke, C.~Richter, F.~Berndt, I.~N. Ioffe,
  A.~A. Granovsky, R.~Mahrwald, N.~P. Ernsting and S.~A. Kovalenko; The Journal
  of Physical Chemistry B \textbf{118} 8756 (2014).

\bibitem{LeScBr12_15296_}
J.~Léonard, I.~Schapiro, J.~Briand, S.~Fusi, R.~R. Paccani, M.~Olivucci and
  S.~Haacke; Chemistry – A European Journal \textbf{18} 15296 (2012).

\bibitem{JiXiLi11_244_}
C.-W. Jiang, R.-H. Xie, F.-L. Li and R.~E. Allen; The Journal of Physical
  Chemistry A \textbf{115} 244 (2011).

\bibitem{FaGrMa12_6367_}
D.~Fazzi, G.~Grancini, M.~Maiuri, D.~Brida, G.~Cerullo and G.~Lanzani; Phys.
  Chem. Chem. Phys. \textbf{14} 6367 (2012).

\bibitem{XiChQu11_926_}
L.~Xiao, Z.~Chen, B.~Qu, J.~Luo, S.~Kong, Q.~Gong and J.~Kido; Advanced
  Materials \textbf{23} 926 (2011).

\bibitem{DuHoLe10_6392_}
L.~Duan, L.~Hou, T.-W. Lee, J.~Qiao, D.~Zhang, G.~Dong, L.~Wang and Y.~Qiu; J.
  Mater. Chem. \textbf{20} 6392 (2010).

\bibitem{Ko96__}
T.~Kobayashi; \emph{J-aggregates}; volume~1; World Scientific (1996).

\bibitem{KueLo11_47_}
O.~K{\"u}hn and S.~Lochbrunner; Semiconductors and Semimetals \textbf{85} 47
  (2011).

\bibitem{SaEiVa13_21_}
S.~K. Saikin, A.~Eisfeld, S.~Valleau and A.~Aspuru-Guzik; Nanophotonics
  \textbf{2} 21 (2013).

\bibitem{DoYaKoe04__}
W.~Domcke, D.~Yarkony and H.~Köppel, editors; \emph{Conical intersections:
  electronic structure, dynamics \& spectroscopy}; World Scientific, Singapore
  (2004).

\bibitem{OlFl15_11428_}
T.~A.~A. Oliver and G.~R. Fleming; The Journal of Physical Chemistry B
  \textbf{119} 11428 (2015).

\bibitem{MaKu11__}
V.~May and O.~K{\"u}hn; \emph{Charge and energy transfer dynamics in molecular
  systems}; John Wiley \& Sons (2011).

\bibitem{WiMo60_872_}
A.~Witkowski and W.~Moffitt; J. Chem. Phys. \textbf{33} 872 (1960).

\bibitem{FuGo61_1059_}
R.~L. Fulton and M.~Gouterman; J. Chem. Phys. \textbf{35} 1059 (1961).

\bibitem{Me63_154_}
R.~E. Merrifield; Radiat. Res. \textbf{20} 154 (1963).

\bibitem{FuGo64_2280_}
R.~L. Fulton and M.~Gouterman; J. Chem. Phys. \textbf{41} 2280 (1964).

\bibitem{BeDaKj02_5810_}
W.~J.~D. Beenken, M.~Dahlbom, P.~Kjellberg and T.~Pullerits; J. Chem. Phys.
  \textbf{117} 5810 (2002).

\bibitem{RoEiDv11_054907_}
J.~Roden, A.~Eisfeld, M.~Dvo\v{r}\'ak, O.~B\"unermann and F.~Stienkemeier;
  Journal of Chemical Physics \textbf{134} 054907 (2011).

\bibitem{HaShCh99_381_}
M.~Hayashi, Y.~J. Shiu, C.~H. Chang, K.~K. Liang, R.~Chang, T.~S. Yang,
  R.~Islampour, J.~Yu and S.~H. Lin; Journal of the Chinese Chemical Society
  \textbf{46} 381 (1999).

\bibitem{EiBrSt05_134103_}
A.~Eisfeld, L.~Braun, W.~T. Strunz, J.~S. Briggs, J.~Beck and V.~Engel; J.
  Chem. Phys. \textbf{122} 134103 (2005).

\bibitem{ScFi84_269_}
P.~O.~J. Scherer and S.~F. Fischer; Chem. Phys. \textbf{86} 269 (1984).

\bibitem{WaEiBr08_044505_}
P.~Walczak, A.~Eisfeld and J.~S. Briggs; J. Chem. Phys. \textbf{128} 044505
  (2008).

\end{thebibliography}
\end{document}